\documentclass[twocolumn,pra,tighten,floatfix,showpacs]{revtex4-1}
\usepackage{graphicx}

\def\<{\langle}
\def\>{\rangle}

\def\be{\begin{equation}}
\def\ee{\end{equation}}

\begin{document}
\preprint{cond-mat} \title{Detecting many-body entanglements in noninteracting ultracold atomic fermi gases}

\author{G. C. Levine*, B. A. Friedman$^\dagger$ and M. J. Bantegui*}

\address{*Department of Physics and Astronomy, Hofstra University,
Hempstead, NY 11549}
\address{$^\dagger$Department of Physics, Sam Houston State University, Huntsville TX 77341}

\date{\today}

\begin{abstract}
We explore the possibility of detecting many-body entanglement using time-of-flight (TOF) momentum correlations in ultracold atomic fermi gases. In analogy to the vacuum correlations responsible for Bekenstein-Hawking black hole entropy, a partitioned atomic gas will exhibit particle-hole correlations responsible for entanglement entropy. The signature of these momentum correlations might be detected by a sensitive TOF type experiment.
\end{abstract}

\maketitle
\section{Introduction} Quantum entanglement seems to provide an important connection between several distinct fields of physics ranging from conformal field theory \cite{cardy_review2}  and topologically ordered phases in quantum field theories \cite{kitaevpreskill} to quantum gravity  \cite{ryu_review}. The central quantity common to these studies is the {\sl entanglement entropy}, computed for a finite subregion of a quantum field theory (QFT) or many body system. If, in a QFT in $d$ spatial dimensions, a distinguished region $A$ of volume $L^{d}$ is formed,  it follows that the degrees of freedom which reside exclusively in the region $A$ will appear to be in a mixed state.  The degree of mixing may be characterized by the entanglement entropy, $S = -{\rm tr}\rho \ln{\rho}$, where  the reduced density matrix $\rho = {\rm tr}_{\notin A}|0\>\<0|$ has been formed by tracing over the degrees of freedom of the ground state, $|0\>$, exterior to the region $A$. 

Entanglement entropy typically obeys an {\sl area law} and is proportional to the area of the bounding surface ($L^{d-1}$), although several variants are possible depending on the underlying particle statistics (fermion or boson) and dimensionality. For many-body fermion systems on a lattice, relevant for the ultracold atom systems discussed here, $S \propto L^{d-1}\log{L}$ for a $d$-dimensional system of $L^d$ sites partitioned as described above.

Entanglement involving several particles has been demonstrated in optical traps with atoms that are specially (quantum mechanically) prepared---a technique critical to quantum computing. However we distinguish this type of entanglement---with a few specially prepared particles---from the entanglement of a many-body system in its ground state, in that the entropy accorded the latter state satisfies an area law and is an intrinsic feature of the many-body ground state, rather than a particular preparation. So far, many-body entanglement entropy has not been measured in any solid state, atomic or electromagnetic system, although an interesting proposal has been made for measuring entropy in a periodically gated quantum point contact \cite{KL_noise}. 

Consider a set of identically prepared systems with a distinguished subsystem $A$. A quantum measurement in region $A$ of any globally conserved quantity will fluctuate in a way that is practically indistinguishable from thermal fluctuations \cite{refael_numberfluct}. For instance, number fluctuations in a partitioned gas of nonrelativistic fermions are indistinguishable from thermal fluctuations at a specific chemical potential. 
\begin{figure}[ht]
\includegraphics[width=7.5cm]{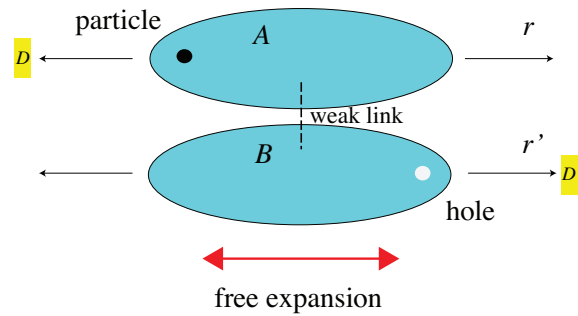}
\caption{\label{fig1} Idealized free expansion experiment to illustrate many-body entanglement in the ground state of an atomic fermion cloud. Two atomic clouds in optical lattices are coupled by a weak link. Analogous to the correlations responsible for Hawing radiation, a particle state in one cloud is correlated with a hole state in the other cloud. Entanglement entropy (the rough analog of Hawking-Bekenstein black hole entropy) is created when the entanglement between particle and hole is severed. These correlations may only be observed if propagation from cloud $B$ to the detector ($D$) for $A$ (and vice versa) is blocked.}
\end{figure}

Our main statement is that the momentum distribution in a series of such measurements  {\sl is} distinguishable from a thermal distribution, and by studying momentum correlations the presence of entanglement might be inferred. In particular, a partitioned fermi gas will exhibit particle-hole correlations across the boundary; these correlations are exactly analogous to the particle-antiparticle correlations responsible for Hawking radiation and Bekenstein-Hawking entropy. In an atomic gas---unlike a black hole---it is possible to look on both sides of the boundary and, in principle, to measure these correlations and infer entanglement. 

The correlations that result from spatially partitioning the fermi gas very much resemble the momentum correlations in a BCS superfluid state---even though the partitioned system is completely noninteracting.  TOF correlations have been very successful in detecting BCS pairing and Mott insulating states in interacting fermi systems \cite{demler}. Here we explore the possibility of using the same detection scheme to look for momentum correlations that are the hallmark of many-body entanglement.

\section{The Schmidt basis for fermions}

Consider two one-dimensional noninteracting fermi atom clouds in separate optical lattices (subsystems $A$ and $B$), connected by a "weak link" at one site (fig. 2). The model hamiltonian is:
\begin{eqnarray}
\label{ham}
H &=& -t\sum_{\< x,y \> \alpha=A,B}{(c^{\dagger}_{x,\alpha}c_{y,\alpha} + c^{\dagger}_{y,\alpha}c_{x,\alpha}})\\
\nonumber&-& w (c^{\dagger}_{0,A}c_{0,B} + c^{\dagger}_{0,B}c_{0,A})
\end{eqnarray}
where $x$ and $y$ are one-dimensional site indices and $A$ and $B$ denote the two identical systems with hopping amplitude $t$ and weak link amplitude, $w$.  Each subsystem consists of $L$ sites and is taken to have fixed boundary conditions at its ends. Furthermore, we will restrict our considerations to the case of $L$ fermions in $2L$ sites. $c_{x}^{\alpha}$ ($c^{\alpha \dagger}_x,$) destroys (creates) fermions at site $x$ in subsytem $\alpha$ and obeys the conventional fermion algebra.  

It is well known that the ground state wave function for such a system may be written in the Schmidt basis in which entanglements between $A$ and $B$ appear in a transparent way. The Schmidt basis is found by diagonalizing the free fermion ground state correlation function matrix
\begin{equation}
\label{corr}
g_{xy} \equiv \< c^\dagger_x c_y\>
\end{equation}
with $x,y$ restricted to subsystem $A$ \cite{peschel_corr_fn}. The eigenvalues $\{n_l\}$, and eigenvectors, $\{A_l(x)\}$, satisfy:
\begin{equation}
\sum_{x \in A}{g_{xy}A_l(y)} = n_l A_l(x)
\end{equation}
A canonical transformation to new set of fermion creation operators, $\{f^{A\dagger}_l\}$, that create fermions in the modes $\{A_l(x)\}$ is made as follows:
\begin{equation}  
\label{unruh}
f^{A\dagger}_l = \sum_{x \in A}{A_l(x) c^{A\dagger}_x} 
\end{equation}
Modes for the $B$ subsystem are formed analogously; in fact if we are restricted to $A$ and $B$ both of size $L$, the modes for subsystem $A$ are also the $\{A_l(x)\}$ but with a coordinate system chosen so that $x=0$ at the boundary of $A$ and $B$. The ground state wavefunction of noninteracting fermions on a lattice may then be written:
\begin{equation}
\label{schmidt} 
|\psi\rangle = \prod_{l=1}^{L}(\sqrt{1-n_l}f_{L-l}^{B\dagger} + \sqrt{n_l} f_l^{A \dagger})|0\rangle_A |0\rangle_B
\end{equation}

It has been known for some time that a system of free fields (bosonic fields or fermionic fields) partitioned into two nonintersecting spatial subsystems has a coherent state wavefunction that resembles the BCS wavefunction of superconductivity.  This wavefunction dates back to earliest explorations quantum fields in curved space and black hole quantum mechanics by Fulling, Parker, Unruh and Hawking \cite{BirrellDavies}.  Klich and others have independently developed this type of wavefunction in connection with entanglement entropy in condensed matter systems \cite{klich_schmidt} and the density matrix renomalization group \cite{dmrg_schmidt}.
\begin{figure}[ht]
\includegraphics[width=7.5cm]{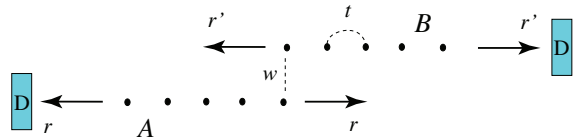}
\caption{\label{fig2} $A$ and $B$ optical lattices with hopping amplitude $t$, coupled by a weak link with amplitude $w$. In idealized TOF experiment, propagation from cloud $A$ is blocked from detector at cloud $B$.}
\end{figure}

The importance of the Schmidt basis for entanglement is that each noninteracting many body fermion state for subsystem $A$ appearing in the wavefunction (a product of $f^{A\dagger}$'s) is correlated with exactly one unique complementary fermion state for subsystem $B$ (a product of $f^{B\dagger}$'s).  Equivalently, the wavefunction may be thought of as a BCS type pairing wavefunction:  the presence of a particle in $A$ in the state $l$ is exactly correlated with a hole in $B$ in the state $L-l$. This wavefunction may also be understood in the weak coupling limit $w << t$ in which subsystems $A$ and $B$ are largely independent fermi seas, each with $L$ states and a pseudo fermi level at $l=L/2$. Deep inside the fermi sea ($l << L/2$), an $f^\dagger_l |0\rangle$ state is approximately equivalent to an unperturbed momentum state $c^\dagger_l |0\rangle$ related to $c^\dagger_x|0\rangle$ in the usual way:
\begin{equation}
\label{unperturbed}
c^\dagger_x = \sqrt{\frac{2}{L+1}} \sum_{k=1}^L{\sin{\frac{k\pi x}{L+1}} c^\dagger_k} = \sum_{k=1}^L{\phi_k(x)
c^\dagger_k} 
\end{equation}

If a TOF type experiment sensitive to the momentum correlations between subsystems $A$ and $B$ could be performed, we would expect fermions in $A$ with pseudo-momentum $l$ to be anticorrelated with fermions in $B$ carrying pseudo-momentum $L-l$. However, unlike the BCS pairing correlations observed in TOF experiments in interacting fermion atomic gases, the momentum correlations in the present noninteracting system are artifacts of the spatial partitioning. 

It is also important to realize that it is only in the Schmidt basis that particles and holes are exactly correlated. In a free expansion experiment particles represented in the Schmidt basis ($f_l^\dagger$) are projected onto particles in the unperturbed momentum basis ($c^\dagger_l$). Close to the pseudo fermi surface, the relationship between the Schmidt basis and the unperturbed momentum basis becomes complicated.

\section{TOF Momentum correlations}

In this manuscript, we illustrate the effect of these particle-hole correlations by an idealized experiment
where two momentum "detectors" are placed at opposite ends of the 1-d sample, shown in figure 2. (In reality, these are optical density measurements). To realize the momentum correlations between $A$ and $B$, the gas must be allowed to freely expand in such a way that particles from subsystem $A$ are blocked from the $B$ detector and vice-versa, as depicted in figure 2.  This might be achieved by suddenly imposing a potential barrier between the two subsystems and imaging the TOF momentum correlations in the conventional way before the wavefunction evolves out of the sudden approximation timescale.    

Following the usual procedure, the spatial intensity-intensity correlation function after a period of free expansion captures the momentum correlations of the original gas. Specifically, we consider the free expansion after time, $t$, of a noninteracting fermi gas in a 1-d optical lattice with spacing, $a_0$. Following Altman et al. \cite{demler}, $d \equiv \hbar t/a_0 m$ is the characteristic size of the wavefunction corresponding to an individual lattice site after free expansion. At time $t$, the continuum creation operator at point $r$, $A^\dagger(r)$, is related to the lattice creation operator as follows:
\begin{equation}
A^\dagger(r) = \sum_{x\in A,B}{w^*_x(r,t)c_x^\dagger}
\end{equation}
$x$ is an integer site index corresponding to subsystem $A$ or $B$ (or both) and $w^*_x(r,t)$ is the freely expanded single lattice site wavefunction, approximately,
\begin{equation}
w^*_x(r,t) \approx \frac{1}{\sqrt{d}}e^{i r x/d}
\end{equation}

The measured particle density---intensity---at point $r$ from atoms originating in subsystem $A$ is then
\begin{equation}
\langle I(r) \rangle_A = \langle A^\dagger (r) A(r) \rangle = \frac{1}{d}\sum_{x,x^\prime \in A}{e^{i(x-x^\prime)\frac{r}{d}}\langle c^\dagger_x c_{x^\prime} \rangle } 
\end{equation}

To form the intensity-intensity correlations function $\langle I(r) I(r^\prime) \rangle_{AB}$ for the partitioned system, we block atom propagation across the $A-B$ boundary during free expansion. The "connected" correlator is then:  
\begin{eqnarray}
\langle I(r) I(r^\prime) \rangle_{AB} - \langle I(r) \rangle_A \langle I(r^\prime) \rangle_B = 
\label{concorrelator}
\\ \nonumber  \frac{1}{d^2} \sum_{x,x^\prime \in A} \sum_{y,y^\prime \in B} e^{i(x-x^\prime)\frac{r}{d}} e^{i(y-y^\prime)\frac{r^\prime}{d}} \langle c^\dagger_x c_{y^\prime} \rangle \langle c_{x^\prime} c^\dagger_y \rangle
\end{eqnarray}
Figure 3 shows a density plot of the connected correlator, eqn. \ref{concorrelator}. Dark shades correspond to negative correlations. The diffuse characteristic of the density-density correlator in fig. 3 is a feature of the finite size of the cluster; for instance, $\langle I(r) \rangle$ for a single 1-d cloud at $T=0$ should resemble the fermi step function, broadened by $2\pi w/L$.  All features in the density-density correlator are rounded by the same factor.  
\begin{figure}[ht]
\includegraphics[width=7.5cm]{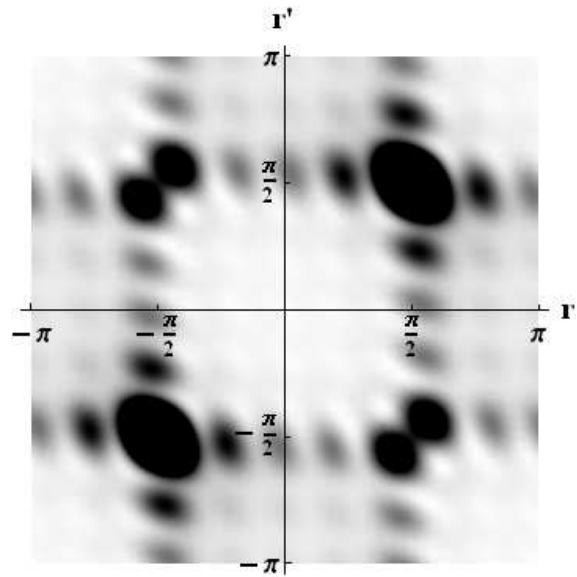}
\caption{\label{fig3} Connected intensity-intensity correlation function simulated for a partitioned ultracold fermi gas ($L = 10$; $w =  t$). The parameter $d$ has been set to unity. Looking at the $(\frac{-\pi}{2},\frac{\pi}{2})$ point in the upper/left quadrant representing correlations at the fermi points of both subsystems, an asymmetry about $r^\prime = \frac{\pi}{2}$ is seen. This asymmetry reflects the anticorrelations (dark) between fermions below the fermi level in $A$ with fermions above the fermi level in $B$}
\end{figure}

The $(\frac{-\pi}{2},\frac{\pi}{2})$ point in the upper left quadrant represents correlations between left moving fermions from $A$ and right moving fermions from $B$, both originating close to the pseudo fermi points of their respective subsystems (that is, with pseudo momentum $l \approx L/2$). The asymmetry under reflection about the line $r^\prime = \pi/2$ is an indication of the anti-correlation between a particle in $A$ in the state $l=L/2+\epsilon$, slightly above the pseudo fermi surface of $A$, with a particle in $B$ in the state $L-l$, slightly below the pseudo fermi surface of $B$. (Specifically, this is the lower left lobe of the two lobes at the $(\frac{-\pi}{2},\frac{\pi}{2})$ point.)

If the ground state corresponded to the unperturbed particle momentum state, $k$, correlated with a hole state $L-k$---as opposed to a correlation between the Schmidt basis particles---the feature seen in fig. 3 would be a diagonal stripe at the $(\frac{-\pi}{2},\frac{\pi}{2})$ point. (That is, the two "lobes" seen at this point would be extended in the diagonal direction.) To understand this figure, consider the canonical relation between the Schmidt basis and unperturbed momentum basis. Combining eqns. (\ref{unruh}) and (\ref{unperturbed}), 
\begin{equation}
\label{bogoliubov}
f^{A\dagger}_l = \sum_{x,k=1}^{L}{A_l(x) \phi_k(x) c^{A\dagger}_k} = \sum_{k=1}^L{P_{lk}c^{A\dagger}_k}
\end{equation}
define the coefficient $P_{lk}$ of the canonical transformation. The sudden approximation, mentioned above, is simply the projection of the ground state wavefunction (\ref{schmidt}) onto the unperturbed momentum basis.

Figure 4 shows a representative set of $P_{lk}$ for $l$ close to the pseudo fermi surface ($l = L/2$) of subsystem $A$. A single Schmidt basis state close to the fermi surface involves an extensive mixture of unperturbed momenta states. The horizontal feature in fig. 3, extending the the right of the $(\frac{-\pi}{2},\frac{\pi}{2})$ point may be understood as the correlation between a single Schmidt particle state, slightly above the pseudo fermi surface in $A$, with a set of unperturbed momentum hole states below the fermi surface in $B$.

\begin{figure}[ht]
\includegraphics[width=7.5cm]{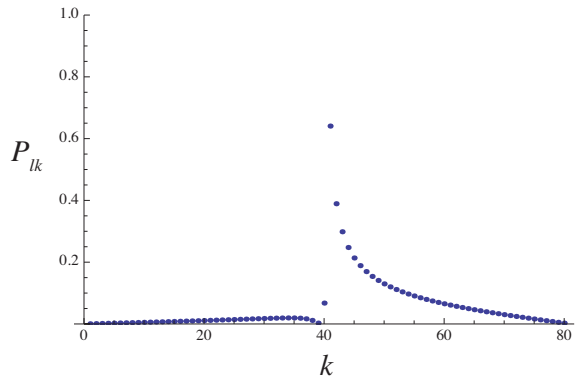}
\caption{\label{fig4} Coefficients $P_{lk}$ for $l$ at the pseudo fermi surface ($l=L/2$) of an $L=80$ lattice. Each Schmidt state close to the fermi surface is an extensive mixture of real momentum states.}
\end{figure}

\section{Number fluctuation and entropy}

As the weak link amplitude $w$ is reduced from the translation invariant case, $w=t$, the connected correlator depicted in fig. 3 does not change qualitatively. However, the integrated weight over the 1st Brillouin zone is approximately proportional to $w^2$, specifically,
\begin{equation}
\label{theweight}
\int_{-\pi}^\pi dr \int_{-\pi}^\pi dr^\prime \langle I(r) I(r^\prime) \rangle_{c} \propto (\frac{w}{t})^2 \log{L}
\end{equation}
where the connected correlator, $\langle I(r) I(r^\prime) \rangle_{c}$, is the first line of eqn. (\ref{concorrelator}). This quantity may be understood as a measure of the number fluctuations in equilibrium of one subsystem \cite{fluct_explanation}. The number fluctuation for fermions in subsystem $A$, $\langle N_A^2 \rangle_c$, may be computed from the wavefunction (\ref{schmidt}) using the canonical transformation (\ref{unruh}):
\begin{equation}
\label{numberfluct}
\langle N_A^2 \rangle_c = \sum_{x,y \in A}{\langle \psi| c_x^\dagger c_x c_y^\dagger c_y| \psi \rangle_c} = \sum_{l=1}^L{n_l(1-n_l)}
\end{equation}

A perturbative estimate (in the parameter $w/t$) of the middle equality in eqn. (\ref{numberfluct}) gives the result (\ref{theweight}). However, the last sum in eqn. (\ref{numberfluct}) may {\sl also} be interpreted as the entropy accorded the fermionic distribution $n_l$ within the Sommerfeld approximation.  Comparing (\ref{concorrelator}) to (\ref{numberfluct}), the integrated weight (\ref{theweight}) in the idealized experiment is proportional to entanglement entropy for noninteracting fermions. Considering the normalization in (\ref{theweight}), an experimental measurement of the integrated weight with sufficient precision to see the exact logarithmic dependence would be difficult; however the dependence upon the weak link amplitude might be more realistic. Simply confirming the particle-hole correlation (the asymmetric feature discussed in connection with fig. 3) would be interesting.

 In a 1-d translation invariant fermion system, it has been rigorously established that the entanglement entropy of a subsystem is proportional to the number fluctuations in that subsystem \cite{KL_noise, refael_numberfluct,klich_schmidt,lehur}. For such a continuum 1-d system, the entanglement entropy is proportional to $\log{L}$ with a universal prefactor that depends upon the central charge of the underlying CFT. 
 
Entanglement entropy when a defect is placed at the boundary of a subsystem has been studied in several ways \cite{levine_imp,peschel_imp1,peschel_imp2} and a remarkable exact solution for the entropy has recently been found by Eisler and Peschel \cite{exact_imp}. When a weak link is placed at the boundary of the two subsystems, the entropy remains logarithmic but, curiously, a feature of the exact solution is that the prefactor is non-analytic about $w=0$. In a sense, the non-analytic behavior at weak coupling is responsible for the behavior in figure 4. At very weak coupling, the entropy is carried by a very few maximally entangled Schmidt pairs. However, despite the weakness of the coupling, each Schmidt state involves an extensive mixture of real momentum states.  Experimental investigation of this feature might also be interesting.
 
\begin{figure}[ht]
\includegraphics[width=7.5cm]{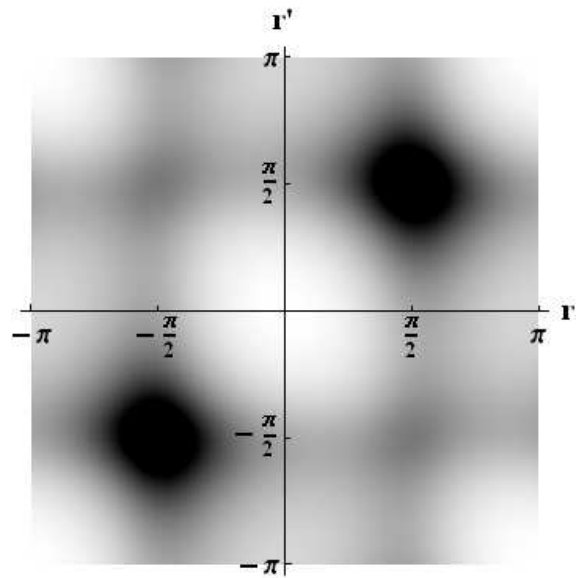}
\caption{\label{fig5} Connected intensity-intensity correlation function simulated for a partitioned ultracold fermi gas ($L = 10$; $w =  t$) at temperature $T=0.5t$. The asymmetry about $r^\prime = \frac{\pi}{2}$ is barely visible and represents an integrated weight that is diminished by approximately $25 \%$ from figure 3.}
\end{figure}

In figure 5 we show the same momentum correlation as in figure 3, but at a temperature of $T=0.5 t$ (as opposed to $T=0$).  As the $(\frac{-\pi}{2},\frac{\pi}{2})$ correlation is only sensitive to entanglement induced number fluctuations---and not thermal number fluctuations---the coherence effect responsible for the particle-hole correlations are quickly lost at finite temperature. The integrated weight about the $(\frac{-\pi}{2},\frac{\pi}{2})$ point is significantly reduced compared to that of figure 3 (by a factor of approximately $0.25$.) Presently there is no analytic description of the temperature dependence of the momentum correlations we have studied. In a single flavor ultracold fermi gas, experimentally accessible temperatures (corresponding to $T \approx 0.1 t$ in our calculation) should exhibit the effect we have described; it does not appear to be necessary to visit an extremely degenerate regime.
 
\section{Conclusion}
 
In conclusion, we have outlined an approach to measuring many body entanglement in the ground state of an ultracold fermi gas in an optical lattice. We have concentrated here on a one-dimensional prototype system, however 2-d or 3-d clouds coupled by a weak link would exhibit a similar effect. The s-wave fermion modes, with respect to the impurity, define a quasi one-dimensional system exactly analogous to the one described in this manuscript \cite{levine_miller}.  Although we have proposed a realization of the experiment involving a potential barrier imposed in the sudden approximation, it is our hope that experimenters will discover a more nuanced approach.

This research was supported by grants from the Department of Energy DE-FG02-08ER64623---Hofstra University Center for Condensed Matter, Research Corporation CC6535 (GL) and NSF Grant no. 0705048 (BF).


\begin{references}
\bibitem{fluct_explanation} The integrated weight is proportional to $\langle N_A N_B \rangle$, but $\langle N_A N_B \rangle = \langle N_A^2 \rangle + \,\, {\rm const}$.
\end{references}
\end{document}